\documentclass[english,aps,pra,showpacs,superscriptaddress,twocolumn,titlepage,longbibliography]{revtex4-1}   
\usepackage{geometry}		% for controlling page margins
\usepackage[above,below]{placeins}	% allows use of \FloatBarrier command to force section barriers
\usepackage{graphicx,color,braket,footnote,subfigure}
\usepackage{amsmath,amsthm,bm}
\usepackage[T1]{fontenc}
\usepackage[utf8]{inputenc}
\usepackage{amssymb}
\newcommand{\bla}{\color{black}}

\usepackage{hyperref}
\hypersetup{colorlinks,linkcolor={red},citecolor={blue},urlcolor={red}}
\usepackage{amsxtra}
\usepackage{lmodern}
\begin{document}
	\title{Temporal self-similarity of quantum dynamical maps as a concept of memorylessness}
	
	\author{Shrikant Utagi}
	\email{shrik@ppisr.res.in}
	\affiliation{Poornaprajna Institute of Scientific Research,
		Sadashivnagar, Bengaluru -  562164,     India}
	\affiliation{Graduate Studies, Manipal Academy of Higher Education, Manipal -576104, India.}
	\author{R. Srikanth}
	\email{srik@poornaprajna.org}
	\affiliation{Poornaprajna Institute of Scientific Research,
		Sadashivnagar, Bengaluru -  562164,     India}
	\author{Subhashish Banerjee.}
	\email{subhashish@iitj.ac.in}
	\affiliation{Interdisciplinary Research Platform- Quantum Information and Computation (IDRP-QIC),
		Indian Institute of Technology, Jodhpur-342037, India.}	
\begin{abstract}
The problem of defining quantum non-Markovianity has proven elusive, with various in-equivalent criteria put forth to address it. The concept of CP-indivisibility and the hierarchy of stronger divisibility criteria going up to P-indivisibility, capture a fundamental aspect of memory in quantum non-Markovianity. In practice, however, there can be a memory-like influence associated with divisible channels in the form of weakening, if not reversing, the effects of decoherence. Arguably, such a facet of memory relates to CP-indivisibility as quantum discord relates to entanglement. We concretize this weaker notion of non-Markovianity by identifying it with deviation from ``temporal self-similarity'', the property of a system dynamics whereby the propagator between two intermediate states is independent of the initial time $t_0$.   We illustrate this idea through examples, and propose a geometric quantification of temporal self-similarity, and show how our approach complements the divisibility-based criterion of quantum non-Markovianity.
\end{abstract}
\flushbottom
\maketitle
\thispagestyle{empty}

\section*{Introduction\label{sec:intro}}
Practical quantum information  processing must inevitably contend with quantum noise  \cite{grabert1988quantum, sbsterngerlach, sbqbm, gpsbsrik}, and in particular, non-Markovian effects in the noise \cite{nonmarkovcrypt, rajagopal2010kraus,george2018thermodynamics, sagnik2019non-Markovian,usha2012,switch}. Classical  Markovianity can be defined in terms of the  divisibility of a process into intermediate transitions, or equivalently in terms of the fall in distinguishability of two states. The quantum adaptation of these ideas to define quantum non-Markovianity is not straightforward, and turns to lead to in-equivalent concepts, essentially because the Kolmogorov hierarchy of classical joint probability distributions cannot be transferred to the quantum case  \cite{vacchini2011markovianity, guarnieri2014quantum,VA17,  breuer2016colloquium, li2018concepts}. 

The classical identification of Markovianity with divisibility leads to a hierarchy of quantum divisibility criteria based on the positivity property of the intermediate map (the propagator of the dynamics between two arbitrary times)
associated with a dynamical process being CP-divisible \cite{RHP10,RHP14} or P-divisible \cite{breuer2009measure}, or an intermediate $k$-divisible \cite{chruscinski2014degree, chruscinski2017detecting}. In Ref. \cite{strasberg2018response} quantum non-Markovian behavior was studied from the perspective of linear response theory. The effort to unify all such definitions into a single framework is important and remains studied by various authors; in this context, cf. \cite{pollock2018non}. It is fair to say that the question of how exactly to characterize memory effects in quantum processes remains a topic of active ongoing research \cite{mazzola2010phenomenological, li2019non, li2020non}.

\begin{figure}[ht!]
	\includegraphics[width=8cm,height=6cm]{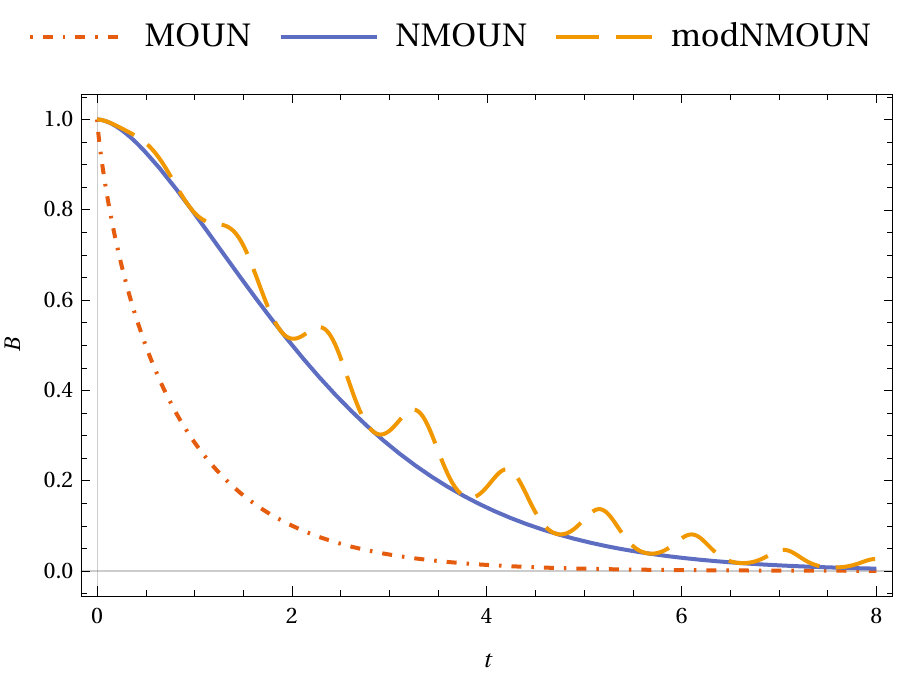}
	\caption{ (Color online)  Distinguishability of two states initially $\ket{+}$ and $\ket{-}$ with respect to time under a Markovian OUN (QDS), non-Markovian OUN (non-QDS but CP-divisible) and modified OUN (CP-indivisible) dephasing channels, quantified by the Holevo bound $B$ (these noisy channels are described below). In the first case, $G=1$ (red, dot-dashed curve), for the second $G=1$ and $g=0.3$ (blue, bold curve), and for the third case, $a=1$ and $r=0.3$, and $k = 1.5$. (orange, dashed curve), cf. Eqs. (\ref{eq:plnoun}), (\ref{eq:modpln}).  The figure illustrates how deviation from QDS, even when it doesn't reverse decoherence, can weaken it.}
	\label{fig:holevoplots} 
\end{figure}

As a rule of thumb, quantum non-Markovianity corresponds to the influence of memory of initial conditions that reverses the effect of quantum decoherence. However, in certain contexts channel memory may manifest itself simply by weakening the effect of quantum decoherence, without actually reversing it.
To illustrate this idea, consider Figure \ref{fig:holevoplots}, which shows how quantum information gets degraded in three dephasing channels: the Holevo bound $B = S(\frac{\rho_1+\rho_2}{2}) - \frac{1}{2}[S(\rho_1)+ S(\rho_2)]$ for a quantum dynamical semigroup channel (bottom-most plot), a CP-indivisible channel (oscillatory plot) and a non-QDS but CP-divisible channel. Although the last mentioned does not manifest recurrence, clearly it involves a memory effect in that it slows down the fall of distinguishability of the two initially orthogonal states ($\ket{+}\bra{+}$ and $\ket{-}\bra{-}$). Here, $S(\rho) \equiv  -\textrm{Tr}[\rho\log(\rho)]$ denotes  the von Neumann entropy.

In the present work, we identify the origin of this memory-like effect in such divisible channels with deviation from ``temporal self-similarity'' (TSS), which roughly refers to the idea of the intermediate map being oblivious of the initial time $t_0$.  Furthermore, we prove that  TSS may be identified with the quantum dynamical semigroup (QDS) \cite{lindblad1976}, the 1-parameter semigroup of maps governing the system dynamics, generated by the time-independent linear map, namely, the Lindbladian $\mathcal{L}$, corresponding to the time-homogeneous master equation $\dot{\rho}(t) = \mathcal{L}[\rho(t)]$ \cite{breuer2002theory,banerjee2018open}. 
Now there has been a traditional background to identifying  quantum Markovianity with QDS \cite{breuer2016colloquium, chruscinski2010non, grabert1988quantum, sbqbm}. This tradition was largely based on two broad physical considerations: (a) it could be justified on grounds that QDS is a reasonable quantum analogue of the classical Chapman-Kolmogorov equation,   (b) it is also favored by considerations of the system-environment interaction, such as a strong coupling with the bath, the correlation times of the environment being very small in relation to the system's relaxation time or allowing for the Born-Markov approximation \cite{grabert1988quantum,sbqbm}.  

However, this historical picture was prior to the emergence of the divisibility-based criteria, and by our revisitation of it is show that it still holds relevance, but in light of a new perspective, which is not explicitly based on parameters related to the system-environment interaction, such as coupling strength, correlation time, etc. (though they play an indirect role, of course), but instead on memory-like effect manifested in the reduced dynamics. In quantum correlation theory, quantum discord  is known to be strictly weaker than quantum entanglement \cite{ollivier2001discord}, and yet of practical relevance \cite{rao2011quantumness}. The discovery of quantum discord was rather surprising in that it highlighted the existence of nonclassicality in the correlation among particles that are classically correlated. We think it is not amiss to consider that the broader concept of quantum non-Markovianity identified here with deviation from TSS stands in relation to the divisibility-based concept of quantum non-Markovanity as quantum discord stands with respect to quantum entanglement. Just as even classically correlated systems can manifest quantum discord, so too even divisible maps can manifest this weaker form of memory. We clarify this point later below.

%\textcolor{blue}{
%In the present work, we will make precise and quantify the broader view of channel memory along the lines of the idea presented in Figure \ref{fig:holevoplots}, namely to regard features of the dynamics that weaken decoherence (but not necessarily reverse it) also as a kind of memory effect. While this constitutes a concept of Markovianity that is stronger than divisibility, it can still be of practical and operational relevance in certain contexts, say understanding how memory prolongs decoherence time-scales, and in the particular contexts of  OUN and PLN.}

%
%For consistency, our approach is shown to integrate with the framework of stronger versions of non-Markovianity, such as CP-indivisibility and P-indivisibility.  We clarify how this can be considered as a weak kind of memory, and provide a geometric quantification of it.

\section*{Temporal self-similarity \label{sec:tss}}
\begin{figure}
	\includegraphics[width=8cm]{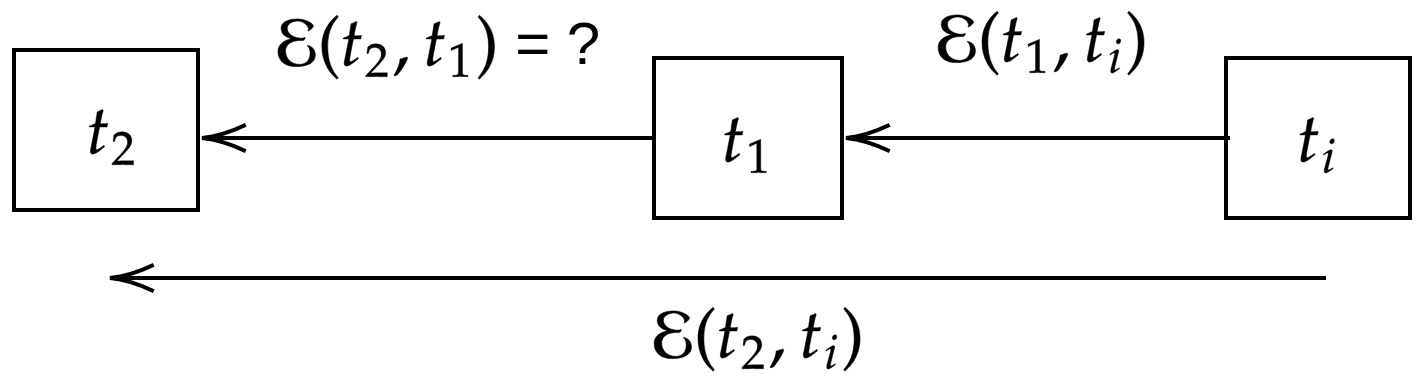}
	\caption{Deciding (non-)Markovianity: Given \textit{the form} $\mathcal{E}(t,t_i)$ (with free variables $t, t_i$) and values $t_1$ and $t_2$, the problem is to compute the intermediate map $\mathcal{E}(t_2,t_1)$. If this computation requires initial data $t_i:= t_0$, then the map is non-Markovian, else it is not. (Equivalently, the original data given to Bob additionally includes $\rho(t_1)$, and his task would be to compute $\rho(t_2)$. Non-Markovianity in this case corresponds to this computation requiring the initial state $\rho$. Given this, $t_0$ can be computed by solving $\rho(t_1)= \mathcal{E}(t_1,t_i)\rho$ for $t_i$, and then $\rho(t_2)$ computed as $\mathcal{E}(t_2,t_0)\mathcal{E}(t_1,t_0)^{-1}\rho(t_1)$. Note that the non-invertibility of $\mathcal{E}(t,t_i)$ does not pose a problem to compute the initial time $t_0$ given $\rho(t_1)$ and $\rho$ \cite[Section IV]{shrikant2018non-Markovian}.)}
	\label{fig:cartoon}
\end{figure}

Suppose Alice obtains states $\rho(t_1)$ and $\rho(t_2)$ by applying CP evolution $\mathcal{E}(t_1,t_i)$ and $\mathcal{E}(t_2,t_i)$, respectively, to a system in initial state $\rho(t_i)$. Here $t_2 > t_1 > t_i$, and $t_i$ is the initial time. She informs Bob the values $t_1, t_2$ and the \textit{form} of the full map $\mathcal{E}(t_f,t_i)$, where $t_f$ denotes the final time of the evolution. The form of the map could be represented as a set of Kraus operators, the Choi matrix, the dynamical map, etc. She asks Bob to compute the intermediate map $\mathcal{E}(t_2,t_1)$ that evolves $\rho(t_1)$ to $\rho(t_2)$. Quite generally $\mathcal{E}(t_2,t_i) = \mathcal{E}(t_2,t_1)\mathcal{E}(t_1,t_i)$. Assuming the invertibility of $\mathcal{E}(t_1,t_i)$, Bob's task is to compute $\mathcal{E}(t_2,t_1) = \mathcal{E}(t_2,t_i)\mathcal{E}(t_1,t_i)^{-1}$ (Figure \ref{fig:cartoon}). 

Clearly, this map is a function of $t_i$ in general. This means that generally only if Alice supplies the value $t_i {:=} t_0$ can Bob compute $\mathcal{E}(t_2,t_1)$. This is evidently a kind of memory effect. (Alternatively, Alice also supplies $\rho(t_1)$ in the beginning, and asks Bob to compute $\rho(t_2)$. In this case, she must reveal the initial state $\rho(t_0)$, reflecting the history dependence of the later state.) On the other hand, if the channel is \textit{temporally self-similar}, i.e., the form of intermediate map is the same as that of the full map, then Bob can simply compute $\mathcal{E}(t_2,t_1)$ by $\mathcal{E}(t_f,t_i)|_{t_f {:=} t_2, t_i {:=} t_1}$. In this case, he doesn't require the $t_0$ information, which gives a notion of memorylessness, and which will be shown to be stronger than CP-divisibility. Thus, a channel $\mathcal{E}$ is \textit{temporally self-similar} when its any intermediate map is oblivious of the initial time, and behaves like a full map in its own right \cite{chandrashekar2007symmetries}. We note that as a valid quantum channel, $\mathcal{E}$ should be completely positive (CP).

Assuming a constant system-bath Hamiltonian, $\mathcal{E}(t,t_i) = \mathcal{E}(t-t_i)$, which essentially follows from the fact that the CP map can be purified to a system-environment unitary $U(t-t_i)$ acting on a product state. Therefore, we have  $\mathcal{E}(t_2,t_1) = \mathcal{E}(f(t_2-t_i,t_1-t_i))$ such that $f(t_2-t_i,t_1-t_i)$ has no dependence of $t_i$ for arbitrary $t_2, t_1$ and $t_i$. Clearly, this holds only if $f(y,x) = f(y-x)$ for $y \ge x$. Setting $r\equiv t_1-t_i$ and $s\equiv t_2-t_1$, we  find that $\mathcal{E}(r+s) = \mathcal{E}(s)\mathcal{E}(r)$, which is the defining composition rule for QDS. Conversely, QDS satisfies temporal self-similarity since the intermediate map under QDS is just $\mathcal{E}(t_2-t_1)$. We thus identify temporal self-similarity with QDS.  

It is worth comparing and contrasting our invocation of initial time memory with that proposed in Ref. \cite{chruscinski2010non}. There,  a non-Markovian quantum evolution (obtained as the reduced dynamics of a time-independent Hamiltonian dynamics defined on the system and an ancilla) is identified with a time-local equation for the dynamical map $\mathcal{E}(t,t_0)$ with a time-homogeneous generator $\mathcal{L}(t-t_0)$ containing a memory of $t_0$. This is indeed analogous to our identification of memory with $t_0$ dependence of the intermediate map. On the other hand, the evolution described by a generalized Lindblad master equation with time-dependent generator $\mathcal{L}(t)$, deviating from QDS but satisfying inhomogeneous composition rule, is taken to be Markovian. Thus, this approach leads to a concept of non-Markovianity stronger than that proposed here. The reason, essentially, has to do with the fact that the initial time dependence  in the case of non-Markovianity occurs in the generator $\mathcal{L}$ via a ``$t-t_0$'' term in the case of Ref. \cite{chruscinski2010non},  whereas it occurs in the intermediate map $\mathcal{E}(t_2,t_1)$ in our case in more general functional forms of dependence on initial time (cf. herebelow Eq. (\ref{eq:choi}) and the discussion below it).

We present a simple illustration of temporal self-similarity as below. Consider the amplitude damping channel (ADC), under which a quantum state $\rho$ evolves to state $\rho^\prime$ via the map $ \mathcal{E}^{\rm AD}[\rho] \rightarrow \sum_j A_j(t) \rho A_j^{\dagger}(t)$, with the Kraus operators $A_j(t)$ given by  $
A_1(t) = \left(
\begin{array}{cc}
1 & 0 \\
0 & \sqrt{1-\lambda(t) } \\
\end{array}
\right)$ and
$A_2(t) = \left(
\begin{array}{cc}
0 & \sqrt{\lambda(t) } \\
0 & 0 \\
\end{array}
\right), $
where
\begin{equation}
\lambda(t) = 1 - e^{-\gamma_0 t}
\label{eq:dampfac}
\end{equation}
is the damping factor and $\gamma_0$ is the vacuum bath interaction parameter \cite{srikanth2008squeezed,omkar2013dissipative}.

Now, suppose that the system evolves starting at $t_i$, going through $t_1$ to $t_2$.
Let the damping factor associated with the full map  $\mathcal{E}^{\rm AD}(t_2,t_i)$  be  $\lambda$, and that with the initial map  $\mathcal{E}^{\rm AD}(t_1,t_i)$  be $\mu$. For  $\mathcal{E}^{\rm AD}(t_2,t_i)$  the Choi matrix \cite{choi1975completely} is found to be \cite{RHP14,kumar2017nonmarkovian}: 
\begin{align}
\chi(t_2,t_i) = \left(
\begin{array}{cccc}
1 & 0 & 0 & \sqrt{1-\lambda } \\
0 & \lambda  & 0 & 0 \\
0 & 0 & 0 & 0 \\
\sqrt{1-\lambda } & 0 & 0 & 1-\lambda  \\
\end{array}
\right),
\label{eq:AD1}
\end{align}
while that for the intermediate evolution map $\mathcal{E}^{\rm AD}(t_2,t_1) = \mathcal{E}^{\rm AD}(t_2,t_i)\mathcal{E}^{\rm AD}(t_1,t_i)^{-1}$ is found to be \cite{kumar2017nonmarkovian}:
\begin{align}
\chi(t_2,t_1) = \left(
\begin{array}{cccc}
1 & 0 & 0 & \sqrt{\frac{1-\lambda }{1-\mu }} \\
0 & \frac{\mu -\lambda }{\mu -1} & 0 & 0 \\
0 & 0 & 0 & 0 \\
\sqrt{\frac{1-\lambda }{1-\mu }} & 0 & 0 & \frac{1-\lambda}{1-\mu} \\
\end{array}
\right).
\label{eq:AD2}
\end{align}
Note that the matrices $\chi(t_2,t_i) $ and $\chi(t_2,t_1)$   are of the \textit{same} form  
provided the functions $1-\lambda$ and $1-\mu$  have  the exponential form $e^{kt}$ for some $k$. In view of Eq. (\ref{eq:dampfac}), this is indeed the case. We thus confirm that amplitude damping is indeed temporally self-similar. 

If the Lindbladian $\mathcal{L}(t)=\mathcal{L}$ is a constant, then  $\mathcal{E}(t_2,t_1) = e^{(t_2-t_0)\mathcal{L}} e^{-(t_1-t_0)\mathcal{L}}= e^{(t_2-t_1)\mathcal{L}} = \mathcal{E}(t_2-t_1)$, i.e., we obtain the self-similar form.  But the converse is not true. This happens essentially when suitable continuity and limit requirements are not met. A simple example here would be the temporally self-similar map $\mathcal{E} = \mathbb{I}$ for $t=0$, and $\mathcal{E}(\rho)  = \sum_j \Pi_j \textrm{Tr}(\rho\Pi_j)$ for $t>0$, where $\Pi_j$ is any complete set of projectors  \cite{li2018concepts}. However, in physically motivated scenarios, we can assume that the channel is continuous, satisfying the limit requirement $\lim_{t\rightarrow 0^+} \mathcal{E} = \mathbb{I}$. Given these assumptions, we can identify self-similarity at the master equation level with the time-independent Lindbladian. Consider the integro-differential time-nonlocal representation of the master equation of the system dynamics in terms of the linear memory kernel map $\mathcal{M}$, which makes this idea of memorylessness clearer. We have $\dot{\rho}(t)=\int_{t_0}^t ds \mathcal{M}_{t-s}\rho(s) = \int_{t_0}^t ds \mathcal{M}_{t-s} [\mathcal{E}(s-t_0)\mathcal{E}^{-1}(t-t_0)]\rho(t) = \int_{t_0}^t ds \mathcal{M}_{t-s} [e^{(s-t)\mathcal{L}}]\rho(t)$. This implies that $\mathcal{M}_{t-s} = \delta(t-s)\mathcal{L}$, where $\delta(t-s)$ is the Dirac delta function, meaning that the dynamics remembers only the present time and has no influence from earlier times.

It is important to stress that this notion of memory  as a dependence on the initial time (or, equivalently, break-down of temporal self-similarity of the map) shows up at the level of maps, and is not obvious at the level of generators. 
If the intermediate map $\mathcal{E}(t_2,t_1)$ is NCP, then it follows that $\mathcal{E}(t_2,t_i)$ is not QDS,  and thus computing $\mathcal{E}(t_2,t_1)$ requires knowledge of $t_i$. Thus, TSS corresponds to a stronger concept of memorylessness than CP-divisibility.  

We may expand on our analogy in the context of quantum correlations, where it is useful to invoke a result due to Sudarshan and coworkers \cite{sudarshan1961stochastic}. It is known that (a subset of) entangled states turn negative under partial transpose, which can serve as a witness of entanglement \cite{horodecki2009quantum}. Discordant states that are separable are necessarily positive under partial transpose. Now, CP-indivisible maps are associated with an intermediate matrix $A(t+\Delta t, t)$ that under an involution operation (cf Appendix) yields the intermediate $B$ map or Choi matrix, which can be negative. That this negativity can be a witness to CP-indivisibility is precisely the RHP criterion \cite{RHP10}. By contrast, non-TSS states that are CP-divisible yield positive matrices under this involution of the intermediate map. Therefore, in the scheme of this analogy, CP-divisibility corresponds to separable states, whilst TSS to non-discordant states. On this basis, we may regard CP-indivisibilty as representing a fundamental non-classical aspect of memory in quantum non-Markovianity, whilst deviation from TSS can include a ``classical-like'' aspect of memory in quantum non-Markovianity.

\section*{Examples: Application to OUN and PLN \label{sec:oupl}}
As a particular instance of divisible noisy channels that manifest this sort of memory effect, we mention two quantum channels, namely Ornstein-Uhlenbeck noise (OUN) and power-law noise (PLN), that were traditionally introduced as non-Markovian based on certain physical arguments, but are Markovian in the CP-divisibility sense. The OUN model was developed in \cite{yu2010entanglement} in the context of the effect of non-Markovian evolution on the dynamics of entanglement. The model used was that of Gaussian noise with a colored auto-correlation function, modeling random frequency fluctuations and has its roots in the modern development of statistical mechanics \cite{uhlenbeck1930theory}. In the limit of infinite noise bandwidth, this reduced to the well-known white noise which is Markovian in nature. PLN is  a  non-Markovian  stationary  noise  process.  The name {\it Power Law} points to the functional relationship between  the  spectral  density  and  the  frequency  of  the  noise.   It  is  a  major  source  of decoherence in solid state quantum information processing devices such as superconducting qubits and has a well-defined Markovian limit \cite{paladino20141}.

The canonical Kraus representation for these channels has the form $\mathcal{E}(\rho) \equiv \sum_{j=I, Z}K_j \rho K_j^\dagger$ with $K_I(t) = \sqrt{\frac{1+p(t)}{2}}I$ and $K_Z(t) = \sqrt{\frac{1-p(t)}{2}}Z$, corresponding to the Choi matrix, $\chi \equiv (\mathcal{E}\otimes \mathbb{I})(\ket{00}+\ket{11})$
given by:
\small
\begin{align}
\chi(t,0) = \begin{pmatrix}
1 & 0 & 0 & p(t) \\
0 & 0 & 0 & 0 \\
0 & 0 & 0 & 0 \\
p(t) & 0 & 0 & 1
\end{pmatrix},
\label{eq:iz}
\end{align}
\normalsize
where $ I $ and $ Z $ are Pauli operators, and
\small
\begin{equation}
p(t)  = \left\{
\begin{array}{ll}
\exp \left(\frac{-G}{2}\left(g^{-1}[\exp (-gt)-1]+t\right)\right)
& \mbox{~case of OUN},\\
\exp \left(-\frac{G t (g t+2)}{2(g t+1)^2}\right)
& \mbox{~case of PLN}.
\end{array} \right.
\label{eq:plnoun} 
\end{equation} 
\normalsize
Here $G$ is the inverse of the effective relaxation time,  while $g$ and $1/g$ are related to the noise band width, for the OUN and PLN noises, respectively.  % both having the dimension of inverse time.

The corresponding master   equation   in   its  canonical   form \cite{andersson2007finding} is  $\dot{\rho} = \gamma(t)(-\rho + Z\rho Z) \label{eq:ME}$ where   $\gamma(t) = -\frac{\dot{p}(t)}{2p(t)} \label{eq:p}$  \cite{shrikant2018non-Markovian} is the  decoherence  rate. It follows from  Eqs. (\ref{eq:plnoun})    
\small
\begin{equation}
\gamma(t) = \left\{
\begin{array}{ll}
\frac{1}{4} G \left(1-e^{-gt}\right)
& \mbox{~case of OUN},\\
\frac{G}{2 (g t+1)^3}
& \mbox{~case of PLN},
\end{array} \right.
\label{eq:2rates} 
\end{equation} 
\normalsize
showing that $\gamma(t)$  remains positive for all $t$ in both these cases.   Thus, neither of them is CP-indivisibe \cite{hall2014canonical}. 

According to the system-environment criterion,  OUN (resp., PLN) have their  Markovian limit by setting  $g$ (resp., $1/g$) $\rightarrow \infty$, in  Eq. (\ref{eq:plnoun}).  In  this limit, we may replace Eq. (\ref{eq:plnoun}) by $p^\ast(t) = e^{-Gt/2}$ (for OUN) and $p^\ast(t) = e^{-Gt}$ (for PLN). The corresponding rate  $\gamma^\ast \equiv -\dot{p}^\ast/(2p^\ast)$ becomes a positive constant, $\frac{G}{4}$ and $\frac{G}{2}$, respectively, in which case the master equation corresponds to the strict (i.e., time-homogeneous) Gorini-Kossakowski-Sudarshan-Lindblad (GKSL) equation \cite{gorini1976,lindblad1976}.  Hence this would correspond to the QDS-limit.
%\begin{equation}
%p^\ast(t) = \left\{
%\begin{array}{ll}
%e^{-Gt/2} & \mbox{~case of OUN},\\
%e^{-Gt} & \mbox{~case of PLN}.
%\end{array} \right.
%\label{eq:plnoun2} 
%\end{equation} 
For the general case, satisfying Eq. (\ref{eq:plnoun}), the decoherence rate  is time-dependent,  implying that the dynamics corresponds to the time-inhomogeneous GKSL  master equation.

The Choi matrix of the intermediate map that evolves the system from time $t_1$ to $t_2$ is given by (cf. the Appendix, where it is derived via the formalism of dynamical maps $A$ and $B$ \cite{sudarshan1961stochastic}): 
\begin{align}
\chi(t_2,t_1) = \begin{pmatrix}
1 & 0 & 0 & \frac{p(t_2)}{p(t_1)} \\
0 & 0 & 0 & 0 \\
0 & 0 & 0 & 0 \\
\frac{p(t_2)}{p(t_1)} & 0 & 0 & 1
\end{pmatrix}.
\label{eq:choi}
\end{align}
From Eq. (\ref{eq:plnoun}), we have for PLN that $p(t_j) = \exp \left(-\frac{G [t_j-t_0] (g [t_j-t_0]+2)}{2(g [t_j-t_0]+1)^2}\right)$. From this, we readily find that the decoherence term $\frac{p(t_2)}{p(t_1)}$ in Eq. (\ref{eq:choi}) does not simplify to a form that is independent of $t_0$.  A similar argument holds for OUN. Thus, even though these two channels are CP-divisible, they carry a memory of the initial time $t_0$, which is required to construct the propagator between any two arbitrary instances.

On the other hand, for the QDS-limit rates, we find $\frac{p(t_2)}{p(t_1)} = e^{-\frac{G}{2}(t_2-t_1)}$ and $e^{-G(t_2-t_1)}$, respectively, for OUN and PLN, showing that the intermediate map is oblivious of $t_0$.

This brings us to the important issue of empirical or practical implication of such weaker-than-CP-indivisible memory-like effect. Two examples may be pointed out. In the OUN model, this effect was made use of \cite{yu2010entanglement} to study its impact on the prolongation of the time to entanglement sudden death (ESD); while in \cite{kumar2017nonmarkovian}, the analogous nature in the OUN and PLN models is shown to counteract decoherence for quantum walks. In both these examples, the weaker-than-CP-indivisible effect acts as a memory resource.

In the context of Figure \ref{fig:holevoplots}, we introduce a non-Markovian noise, which is a modification of non-Markovian OUN inspired by the random-telegraph noise (RTN), by introducing the time-dependent mixing parameter given by
\begin{align}
\tilde{p}(t) = \exp \left[\frac{-a }{2} \left(\frac{e^{-rt}-1}{r}+t\right) \left(\frac{\sin ^2(\frac{t}{r})}{k}+\cos ^2(\frac{t}{r}) \right)\right],
\label{eq:modpln}
\end{align}  and call it modified NMOUN. Here $k$ is some real number. Modified OUN (or, modOUN) has a QDS limit when $r \rightarrow \infty$. It is CP-indivisible if $k \ne 1$.

\section*{Quantifying deviation from temporal self-similarity \label{sec:quant}}

The above considerations suggest that the non-Markovianity in this weaker sense may be geometrically quantified by the minimum distance of an evolution $\mathcal{E}$ from a QDS form either at the level of maps or of generators. In the discrete-time case, only the former is possible. In practice, this approach can be computationally complicated to realize, given the non-convex nature of set of CP-divisible (including QDS) maps \cite{chruscinski2010non}. Typically, for physically well motivated channels, we may assume that the channel satisfies time-continuity and the above-mentioned  limit assumption. Accordingly, as one possibility to quantify non-Markovianity in this sense, one may consider minimizing the distance $\Vert \mathcal{E}(t) - e^{\mathcal{L}^\ast t}\Vert$  for arbitrary maps, where $\mathcal{L}^\ast$ is a time-independent Lindblad generator. But here again one must contend with the non-convexity issue, noted above.

To circumvent this difficulty, we propose here to realize this geometric measure at the level of generators,  instead of maps. We require to define a suitable measure that quantifies $\vert \mathcal{L}(t) -  \mathcal{L}^\ast\vert$. To this end, we consider the infinitesimal intermediate map $\delta\mathcal{E}$ of the given channel $\mathcal{E}$, evolving the system of Hilbert space dimension $d$ from time $t$ to time  $t+dt$. We have $(\delta\mathcal{E})\rho(t) = \mathcal{T}\exp\left(\int_t^{t+dt}\mathcal{L}(s)ds\right)\rho(t)=  (1+\mathcal{L}(t)dt)\rho(t).$
%\begin{eqnarray}
%(\delta\mathcal{E})\rho(t) &=&  \mathcal{T}\exp\left(\int_t^{t+dt}\mathcal{L}(s)ds\right)\rho(t) \nonumber\\
%&=& (1+\mathcal{L}(t)dt)\rho(t).
%\label{eq:partial}
%\end{eqnarray}
By the Choi-Jamiolkowski isomorphism,  for any map $\mathcal{E}$ acting on a $d$-dimensional system, we can associate the unique $d^2 \times d^2$ Choi matrix $\hat{\mathcal{E}}(t) \equiv d(\mathcal{E} \otimes \mathbb{I})\ket{\Phi^+}\bra{\Phi^+}$, where $\ket{\Phi^+} \equiv d^{-1/2}\sum_j \ket{j,j}$ is the maximally entangled state. The Choi matrix of the infinitesimal intermediate map defined above is thus $(d\ket{\Phi^+}\bra{\Phi^+} + \hat{\mathcal{L}}(t)dt)$. Here, the uniqueness means that the Choi matrix is insensitive to the unitary freedom in the representation of the generator, and in particular, is independent of whether the generator has been represented in its canonical form. Let $\delta\mathcal{E}^\ast(t)$ be the infinitesimal intermediate map of a QDS channel, and $\mathcal{L}^\ast$ the corresponding generator. Then  the difference between the two infinitesimal intermediate maps is $\Delta L \equiv  \delta\mathcal{E}(t) - \delta\mathcal{E}^\ast(t) = (\mathcal{L}(t) -  \mathcal{L}^\ast)dt$, from which we have:
\begin{equation}
\zeta = \min_{\mathcal{L}^\ast} \frac{1}{T} \int_0^T \Vert \hat{\mathcal{L}}(t) -  \hat{\mathcal{L}}^\ast\Vert dt,
\label{eq:fqds}
\end{equation}
where $\Vert A \Vert = {\rm Tr}\sqrt{AA^{\dagger}}$ is the trace norm of a matrix $A$. 

This measure of non-Markovianity has the following desirable properties. There is an inherent normalization, whereby $\zeta=0$ iff the channel is QDS and $\zeta>0$ otherwise. The measure is easily computable since the general numerical optimization problem of Eq. (\ref{eq:fqds}) can be implemented in $d^2-1$ dimensions efficiently. The Lindbladians can be represented in their canonical forms, and those with positive Lindblad terms will form a convex set. Furthermore, for specific examples we consider, the minimization over the Lindbladian reduces to the problem of minimizing over a single parameter. The measure respects continuity, which we illustrate with a few examples. It is basis-independent, and has the operational meaning of non-Markovianity as deviating from the property of self-similarity and thus requiring memory of the initial time $t_0$.

A similar geometric approach to quantifying non-Markovianity is considered in \cite{wolf2008assessing}, namely minimizing the distance $\Vert \mathcal{E}(t=1) - e^{\mathcal{L}^\ast}\Vert$  for arbitrary maps. They consider the distance of a map (snapshot of an evolution), whereas we consider the evolution over a finite period of time. The problem with considering a snapshot is that there can be infinite number of evolutions passing through the map at that instant. For example, the map Eq. (\ref{eq:iz}), with a fixed $p(t=T)$, has an infinite number of ways to assign values to the pair $(g,G)$.  

The measure Eq. (\ref{eq:fqds}) is now applied to a number of well-known channels. Immediately below, we consider the CP-divisible channels of OUN and PLN to quantify their non-Markovianity in the present approach.  For dephasing channels, we find $\mathcal{L}\rho = \gamma(-\rho+ Z \rho Z)$.  Thus, $\hat{\mathcal{L}}(t) -  \hat{\mathcal{L}}^\ast =  (\gamma^\ast -\gamma)(\ket{\Phi^+}\bra{\Phi^+} - \ket{\Phi^-}\bra{\Phi^-})$, 
where $\ket{\Phi^-}$ is the Bell state with even parity and negative phase. 

\begin{figure}
	\includegraphics[width=8cm, height=6cm]{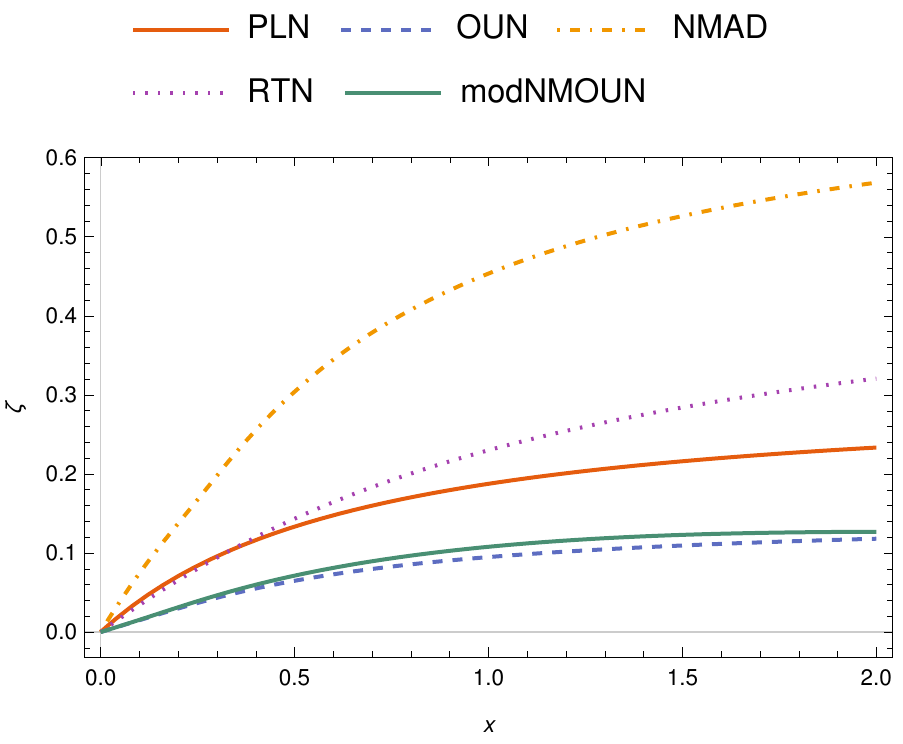}
	\caption{\small (Color online) Upper bound on the non-Markovianity measure $\zeta$ for CP-divisible and CP-indivisible channels, with $T\equiv 1$ (Eq. (\ref{eq:fqds})). Here, $x \equiv g^{-1}$ for OUN, NMAD and RTN and $x \equiv g$ for PLN. The value of parameters used are: $G=0.6$ for both PLN (bold, red curve) and OUN (dashed, blue), and $\gamma_0=0.3$ for NM-AD (dot-dashed, orange) and $\gamma_0=0.6$ for RTN (dotted, purple). The curve (bold, dark green) is an upper bound on the non-Markovianity of modified NMOUN with $k=1.5$ and $a=0.6$, and $x \equiv r^{-1}$.}
	\label{fig:plots}
\end{figure}

For example, for OUN, it follows from their canonical master equation that $\zeta = \min_{c}\frac{1}{T}\int_0^T  \left| \frac{1}{4} G \left(1-e^{-gt}\right) - c\right| dt \label{eq:zetax}$
where $c$ is the  rate of a QDS dephasing channel. The minimization may be determined numerically and would in general depend on $T$. For OUN and PLN, it is simpler to consider an estimate of $c$ to be the family's natural QDS limit, as may be found from the decoherence rates and $p^\ast$, given below Eq. (\ref{eq:2rates}). The corresponding plot (Figure \ref{fig:plots}) may be considered as giving an upper bound on the weak memory effect. Similarly for the modified NMOUN case, Eq. (\ref{eq:modpln}), for which the corresponding plot appears in Figure (\ref{fig:plots}). 

\section*{Application to stronger manifestations of non-Markovianity \label{sec:xt}}
Although the quantity $\zeta$ has been motivated to explore the idea of memory weaker than CP-indivisibility, one would expect it to be applicable to such stronger manifestations of non-Markovianity,  since in those cases, the deviation from the QDS form would be greater.  For completeness, we shall consider a few representative examples, one that is P-divisible but CP-indivisible, another that is P-indivisible, involving both unital and non-unital maps. We present detailed analyses of these in the following.

\subsection*{Eternal non-Markovian channels \label{sec:eternal}} An example of a model noise which is non-Markovian in the sense of CP-indivisibility but BLP Markovian \cite{breuer2009measure} (i.e., P-divisible), is the interesting model called ``Eternally non-Markovian'' (ENM) Pauli channel, proposed in \cite{hall2014canonical}, with the decay rate $\gamma_3(t)$ being negative for all $t>0$, whence the name `eternal'. The canonical form of master equation for the evolution of qubit under this noise is the dephasing channel
$
\dot{\rho} = \sum_{j=1}^{3} \gamma_j(p) (\sigma_j \rho \sigma_j^\dagger - \rho),
$ with $\gamma_1 = \gamma_2 = 1 \, {\rm and } \, \gamma_3(t) = - {\rm tanh}(t)$. 

The measure Eq. (\ref{eq:fqds} \bla), \bla in this case is found to be 
$\zeta = \min_c \frac{1}{T}\int_0^T |-\tanh(t) - c|dt$, for which optimal QDS channel is clearly the dephasing channel with $\gamma_1=\gamma_2=1$ and $\gamma_3=c$, and $c=0$. Setting $T\equiv1$, the degree of non-Markovianity here is $\int_0^1 \tanh(t) dt = \log[\cosh(1)] \approx 0.433$.

\subsection*{Random telegraph noise (RTN): P-indivisible dephasing \label{sec:rtn}}
As our next example, we consider random telegraph noise (RTN), which is a very well studied pure dephasing process, known to be non-Markovian according to information back-flow and CP-divisibility criteria \cite{daffer2004depolarizing, kumar2017nonmarkovian}.  The Kraus operators characterizing this process has a functional form similar to that of PLN and OUN with the function $p(t)$ in Eq. \ref{eq:plnoun} having the form $
p(t) = \exp\{-g t\}\left(\cos(g\omega t) + \frac{\sin(g\omega t)}{\omega}\right),
$ with $\omega = \sqrt{(\frac{2 \gamma_0}{g})^2 - 1}$. Here $g$ is the spectral band width, which is the inverse of environmental correlation time scale $\tau$, and $\gamma_0$ defines the coupling strength between the system and the environment. The decay rate is found to be $\gamma(t) = \frac{2 \gamma_0^2}{g \left( 1+  \sqrt{\frac{4 \gamma_0^2}{g^2}-1} \cot \left(g t \sqrt{\frac{4 \gamma_0^2}{g^2}-1}\right) \right)}$, 
%\begin{align}
%\gamma(t) = \frac{2 \gamma_0^2}{g \left( 1+  \sqrt{\frac{4 \gamma_0^2}{g^2}-1} \cot \left(g t \sqrt{\frac{4 \gamma_0^2}{g^2}-1}\right) \right)},
%\label{eq:rtngamma}
%\end{align} 
which vanishes in the limit $g \gg 2\gamma_0$ and represents the QDS limit of the family. 
In the limit $g > 2\gamma_0$, the process is described by a master equation with time-dependent all-time-positive decay rate $\gamma(t)$, and hence  CP- and P-divisible. This would correspond to Markovian behavior from the  divisibility perspective,  but from the present point of view would be non-Markovian due to deviation from QDS. In the limit $g < 2\gamma_0$, the process becomes non-Markovian in the sense of CP-divisibility and $\gamma(t)$ oscillates between negative and positive values, giving rise to intervals of breakdown of CP-divisibility and even P-divisibility. As $g \rightarrow 0$, the noise becomes more  colored, hence non-Markovian. Setting $\gamma^\ast = 0$ for this noise, an upper bound for the measure of non-Markovianity Eq. (\ref{eq:fqds}), for channels in this family parametrized by $x\equiv\frac{1}{g}$ is depicted in Figure (\ref{fig:plots}).
%A plot of self-similarity as a function of the parameter $g$ for RTN. 

\subsection*{Non-Markovian amplitude damping \label{sec:nmad}}

%Consider the amplitude damping channel (ADC), as discussed above in Appendix \ref{sec:adTSS}.

%under which a quantum state $\rho$ evolves to state $\rho^\prime$ via the map $ \mathcal{E}^{\rm AD}[\rho] \rightarrow \sum_j A_j(t) \rho A_j^{\dagger}(t)$, with the Kraus operators $A_j(t)$ given by  $
%A_1(t) = \left(
%\begin{array}{cc}
%1 & 0 \\
%0 & \sqrt{1-\lambda(t) } \\
%\end{array}
%\right)$ and
%$A_2(t) = \left(
%\begin{array}{cc}
%0 & \sqrt{\lambda(t) } \\
%0 & 0 \\
%\end{array}
%\right), $
%where $\lambda(t) = 1 - e^{-\gamma_0 t}$
%is the damping factor and $\gamma_0$ is the vacuum bath interaction parameter \cite{srikanth2008squeezed, omkar2013dissipative}.

The above method, considered so far for unital channels, can straightforwardly be extended to the non-unital case. 
As a specific example, we may consider the non-Markovian (P-indivisible) amplitude damping (AD)  as an example of non-unital channel, the NM-AD channel. 

Consider a qubit interacting dissipatively with a bath of harmonic oscillators, whose spectral density is given by the Lorentzian 
$$
I(\omega) = \gamma_0 g^2(2 \pi (\omega_0 + \Delta - \omega)^2 + g^2)^{-1},
$$ 
where $ g $ is  the  width of the spectral density function, centered at a frequency detuned from the
atomic frequency $ \omega_0 $ by an amount $ \Delta $, and the rate $ \gamma_0 $
quantifies the strength of the system-environment coupling. If we assume $\Delta = 0$ (no detuning) i.e., when the qubit is in \textit{resonance} with the central frequency of the bath, then 
the GKSL-like time-dependent master equation, with the rotating wave approximation, is given by 
\begin{align}
\frac{d \rho_s(t)}{dt} =
%& - \frac{i S(t)}{4} [\sigma_z, \rho_s(t)] \nonumber \\&+ 
\gamma(t) [\sigma_- \rho_s(t) \sigma_+ - \frac{1}{2}\{\sigma_+ \sigma_-, \rho_s(t)\}], 
\label{eq:mastereq}
\end{align} 
where %, $S(t) = - 2 \Im[\frac{\dot{G}(t)}{G(t)}] $ is the time-dependent Lamb shift and 
$\gamma(t) = - 2 \Re[\frac{\dot{G}(t)}{G(t)}] $ is the time-dependent decoherence rate, and $G$ is the decoherence function given by
\begin{align}
G(t) = e^{- \frac{g t}{2}} \left( \frac{g }{l} \sinh \left[\frac{l t}{2}\right] + \cosh \left[\frac{l t}{2}\right] \right),
\label{eq:G}
\end{align}
with $ l = \sqrt{g^2 - 2 \gamma_0 g} $ and $\sigma_{\pm}$ are the standard atomic raising (lowering) operators.  This is the time-dependent AD channel and the decay rate is given by 
\begin{align}
\gamma(t) &= - \frac{2}{|G(t)|} \frac{d |G(t)|}{dt} \nonumber \\ &= 2 \Re\left(\frac{\gamma_0 }{\sqrt{1-\frac{2\gamma_0}{g}} \coth \left(\frac{1}{2} gt \sqrt{1-\frac{2 \gamma_0}{g}}\right)+1}\right).
\label{eq:NMADgamma}
\end{align}
Now, the expression Eq. (\ref{eq:dampfac}) for the damping factor of the ADC is replaced by $ \lambda(t) = 1 - |G(t)|^2$.\cite{breuer2016colloquium}  In the limit $g < 2\gamma_0$, the decay rate (\ref{eq:NMADgamma}) oscillates, and becomes negative for certain intervals giving rise to non-Markovian evolution. In the limit $g > 2 \gamma_0$, the dynamics is time-dependent Markovian. (The point $g=2\gamma_0$, however, corresponds to a point at which the time-local master equation lacks a perturbation expansion.) One readily sees that in the limit $ g \gg 2 \gamma_0 $, the decay rate $\gamma(t) = \gamma_0$, i.e., it becomes time-\textit{in}dependent, corresponding to a QDS evolution,  the standard AD channel.

The measure Eq. (\ref{eq:fqds}) is found to be $\zeta = \min_{\gamma^\ast} \frac{1}{T}\int_0^T  |\gamma(t) - \gamma^\ast| (1 + \sqrt{2}) dt
\label{eq:zetax2}$.  Once again, an upper bound of non-Markovianity parameter $\zeta$,  can be obtained by choosing $\gamma^\ast$ to the QDS limit of the family of non-Markovian AD channels. As in the dephasing case, we find that the general optimization of the measure reduces to minimizing over a single parameter $\gamma^\ast$, which can be done numerically and is depicted in Figure (\ref{fig:plots}). The measure Eq. (\ref{eq:fqds})  can be suitably adapted through a renormalization procedure to handle the scenario wherein the generator has singularities, cf. \cite{shrikant2020quasi}. 

% \subsection{Eternally non-Markovian (ENM) dephasing channel: CP-indivisible but P-divisible \label{sec:ENM_Pauli}}

\section*{Discussion and conclusion}

%\section*{Conclusions and Discussions \label{sec:conclu}}
Over the last decade, the concept of what constitutes quantum non-Markovianity has been debated and studied in depth. Here we pointed out that even when the system's dynamical map is divisible, there can be a kind of memory of the initial time $t_0$ encoded into the form of the intermediate map, causing the dynamics to deviate from ``temporal self-similarity'' (TSS). Operationally, this memory effect may simply show up as a mitigation of decoherence, rather than a reversal of decoherence. We identify this concept of memorylessness with the quantum dynamical semigroup (QDS), and quantify it, providing a number of examples. We argue that this weaker concept of quantum non-Markovianity stands in relation to the standard divisibility  based definition of quantum Markovianity, as quantum discord stands with respect to entanglement. Of practical importance is that this measure would be applicable in scenarios where the map is divisible, and yet a memory-like effect is seen to counter decoherence caused, for e.g., by modifying the frequency spectrum of the interacting bath or broadening the memory kernel of the dynamics. To illustrate this, we studied various examples, among them the well known channels of OUN and PLN.

A number of new directions are opened up by the present work.  In \cite{budini2018maximally} it is shown how maximally non-Markovian system dynamics can arise without backflow of information from the system to the environment modelled via classical degrees of freedom. While QDS was historically motivated, but not defined, by considerations of the system-bath interaction, another concept of Markovianity that explicitly relies on such considerations is the quantum regression formula (QRF), which is based on the system's two-point or higher-order correlation functions \cite{swain1981master}. Its relation to divisibility-based criteria for non-Markovianity have been studied by various authors  \cite{li2018concepts,guarnieri2014quantum}. In light of these observations, one might ask what the most general system-environment correlations are that lead to deviation from TSS, and also how to reconcile the two different (but related) viewpoints on non-Markovianity, based on the system dynamics vs that based on knowledge of the system-environment interaction and correlations.

%\begin{figure}
%	\begin{center}
%		\begin{tikzpicture}[scale=1.25]
%		%\draw [cyan,fill] (-2.25,-2.25) rectangle (2.25,2.25);
%		\draw [black] (-2.35,-2.35) rectangle (2.35,2.35);
%		\draw [dashed, red] (0,0) circle [radius = 2.1];
%		\draw[dashed, red](0,0) circle [radius = 1.6];
%		\draw [green,fill] (0,0) circle [radius=1.2];
%		\draw [white,fill] (-0.5,-0.5) rectangle (0.5,0.5);
%		\node at (0,0) {\bf  QDS};
%		\node at (-0.1,0.85) {\textbf{\bf  CP-div}};
%		\node at (0,1.8) {\textbf{\bf k-div}};
%		\node at (1.35,2.15) {\textbf{\bf P-div}};
%		%\node at (1.5,1.85) {\textbf{ \tiny P-div}};
%		\end{tikzpicture}
%	\end{center}
%	\caption{Hierarchy of containments among sets of dynamical processes: those that satisfy temporal self-similarity, or equivalently, the QDS criterion  (\textbf{QDS}); those that are CP-divisible  (\textbf{CP-div}); and those that are P-divisible (\textbf{P-div}).   In the regime where PLN and OUN are said to be non-Markovian in the statistical mechanical sense of coloring of the memory kernel, they fall in $\textbf{CP-div}/{\bf QDS}$.  Examples  of processes  falling in set  ${\bf P-div}/{\bf    CP-div}$    are ENM and those  discussed  in \cite{chruscinski2011divisiblity, liu2013nonunital}. The interim (representing a succession of dashed circles) region is one where a hierarchy of definitions of Markovianity may be given in the sense of $k$-divisibility ({\bf $\bf k$-div}) \cite{chruscinski2014degree}. \bla }
%	\label{fig:heirarchy}
%\end{figure}

In \cite{bhattacharya2018resource}, the authors consider a resource theory of non-Markovianity wherein the Choi matrices corresponding to small-time divisible maps constitute the free states. Our work suggests the possibility of constructing a more relaxed resource theory of small-time maps, wherein resourceful states would correspond to maps of processes deviating from temporal self-similarity. Earlier, we noted the instances where the memory-like signature of the OUN model, as characterized here, helps to prolong the time to ESD \cite{yu2010entanglement}  and also counteracts decoherence in quantum walk \cite{kumar2017nonmarkovian}. This should pave the way for more such examples which could have an impact on broadening the scope of memory in quantum phenomena.

In \cite{chruscinski2014degree},  studying the positivity of the propagator between two arbitrary times in an extended Hilbert space (divisibility), the authors defined a non-Markovianity degree, as the analogue of Schmidt number in
quantum entanglement, such that the analogue of maximally entangled states are maximally non-Markovian quantum dynamics. It would be interesting to study how channels weaker than CP-indivisible fit into this hierarchy, in particular, whether they could be considered as the analogues of separable states with non-vanishing quantum discord.

\appendix
\section*{Appendix}
\section*{Sudarshan $A$ and $B$ matrices for PLN-OUN \label{sec:ABmatrices}} The intermediate dynamics can be studied by way of the dynamical map $A$ introduced by Sudarshan et al \cite{sudarshan1961stochastic}. The map $A$ represents the noise superoperator as a $d^2 \times d^2$ matrix on a vector, obtained by vectorizing the density operator, i.e., $\vec{\rho}^\prime = A(t_1,t_0)\cdot \vec{\rho}$. Thus,  $\rho_{j^\prime k^\prime}^\prime = \sum_{j,k} A_{j^\prime k^\prime;jk} \rho_{jk}$. Here $d$ is the system's Hilbert space dimension. Given a qubit density operator $\begin{bmatrix}
\rho_{00} & \rho_{01} \\ \rho_{10} & \rho_{11} 
\end{bmatrix}$, the channel (4) can be represented as:
\begin{equation}
\begin{pmatrix}
\rho_{00}^\prime  \\ \rho_{01}^\prime \\ \rho_{10}^\prime \\ \rho_{11}^\prime 
\end{pmatrix}
= \begin{pmatrix}
1 & 0 & 0 & 0 \\ 0 & p(t) & 0&0\\0&0&p(t)&0\\0&0&0&1
\end{pmatrix}
\begin{pmatrix}
\rho_{00}  \\ \rho_{01} \\ \rho_{10} \\ \rho_{11}
\end{pmatrix},
\label{eq:dynmap}
\end{equation}
where $p(t)$ is given by the corresponding probabilities $p^\ast(t)$ when temporal self-similarity holds, and by Eq. (5) in the general case. 

Unlike the Choi matrix \cite{choi1975completely}, the map $A$ can be composed directly by matrix multiplication: thus $\vec{\rho}(t_2) = A(t_2,t_1)\cdot A(t_1,t_0)\vec{\rho}(t_0)$. The intermediate dynamical map $A(t_2,t_1)$ can thus be directly computed as $A(t_2,t_1) = A(t_2,t_0)\cdot A(t_1,t_0)^{-1}$, where the inverse is the matrix inverse and assumed to be non-singular. For Eq. (\ref{eq:dynmap}), one readily finds that
\begin{equation}
A(t_2,t_1) = \begin{pmatrix}
1 & 0 & 0 & 0 \\ 0 & \frac{p(t_2)}{p(t_1)} & 0&0\\0&0&\frac{p(t_2)}{p(t_1)}&0\\0&0&0&1
\end{pmatrix}.
\label{eq:dephA}
\end{equation}
By re-shuffling the terms of the $A$-matrix according to an ``involution map'' given by $B_{j^\prime j;k^\prime k} = A_{j^\prime k^\prime;j k}$ \cite{sudarshan1961stochastic}, from Eq. (\ref{eq:dephA}), one obtains the corresponding $B$-matrix, which is just the Choi matrix.

\section*{Acknowledgements}

SU and RS acknowledge financial  support from the Department of Science and Technology (DST), India,   provided   through  the  project EMR/2016/004019. RS and SB acknowledges support from Interdisciplinary Cyber Physical Systems (ICPS) programme of the Department of Science and Technology (DST), India through Grants No.: DST/ICPS/QuST/Theme-1/2019/14 (RS), DST/ICPS/QuST/Theme-1/2019/6 and DST/ICPS/QuST/Theme-1/2019/13. SB also acknowledges support from Interdisciplinary Research Platform- Quantum Information and Computation (IDRP-QIC) at IIT Jodhpur.

%\section*{Author contributions statement}

%RS and SB developed the basic theoretical ideas. SU made further contributions to this and also performed all the calculations.

%\section*{Additional information}

%To include, in this order: \textbf{Accession codes} (where applicable); 
%\textbf{Competing interests}: The authors declare no competing interests.

%The corresponding author is responsible for submitting a \href{http://www.nature.com/srep/policies/index.html#competing}{competing interests statement} on behalf of all authors of the paper. This statement must be included in the submitted article file.

%\bibliographystyle{unsrt}
\bibliography{plnoun}

\begin{thebibliography}{10}

\bibitem{grabert1988quantum}
Hermann Grabert, Peter Schramm, and Gert-Ludwig Ingold.
\newblock Quantum brownian motion: The functional integral approach.
\newblock {\em Physics Reports}, 168(3):115 -- 207, 1988.

\bibitem{sbsterngerlach}
Subhashish Banerjee and R.~Ghosh.
\newblock Quantum theory of a stern-gerlach system in contact with a linearly
  dissipative environment.
\newblock {\em Phys. Rev. A}, 62:042105, Sep 2000.

\bibitem{sbqbm}
Subhashish Banerjee and R.~Ghosh.
\newblock General quantum brownian motion with initially correlated and
  nonlinearly coupled environment.
\newblock {\em Phys. Rev. E}, 67:056120, May 2003.

\bibitem{gpsbsrik}
S.~Banerjee and R.~Srikanth.
\newblock Geometric phase of a qubit interacting with a squeezed-thermal bath.
\newblock {\em The European Physical Journal D}, 46(2):335--344, Feb 2008.

\bibitem{nonmarkovcrypt}
Kishore Thapliyal, Anirban Pathak, and Subhashish Banerjee.
\newblock Quantum cryptography over non-{M}arkovian channels.
\newblock {\em Quantum Information Processing}, 16(5):115, Mar 2017.

\bibitem{rajagopal2010kraus}
AK~Rajagopal, AR~Usha Devi, and RW~Rendell.
\newblock Kraus representation of quantum evolution and fidelity as
  manifestations of {M}arkovian and non-{M}arkovian forms.
\newblock {\em Phys. Rev. A}, 82(4):042107, 2010.

\bibitem{george2018thermodynamics}
George Thomas, Nana Siddharth, Subhashish Banerjee, and Sibasish Ghosh.
\newblock Thermodynamics of non-{M}arkovian reservoirs and heat engines.
\newblock {\em Phys. Rev. E}, 97:062108, Jun 2018.

\bibitem{sagnik2019non-Markovian}
Dipanjan Mandal Sandeep K.~Goyal Sagnik~Chakraborty, Arindam~Mallick and
  Sibasish Ghosh.
\newblock Non-{M}arkovianity of qubit evolution under the action of spin
  environment.
\newblock {\em Scientific Reports}, 9(2987), 2019.

\bibitem{usha2012}
A.~R.~Usha Devi, A.~K. Rajagopal, Sudha, and R.~W. Rendell.
\newblock {\em J. Quantum Inform. Sci.}, 2:47, 2012.

\bibitem{switch}
N.~Srinatha, S.~Omkar, R.~Srikanth, Subhashish Banerjee, and Anirban Pathak.
\newblock The quantum cryptographic switch.
\newblock {\em Quantum Information Processing}, 13(1):59--70, Jan 2014.

\bibitem{vacchini2011markovianity}
Bassano Vacchini, Andrea Smirne, Elsi-Mari Laine, Jyrki Piilo, and Heinz-Peter
  Breuer.
\newblock {M}arkovianity and non-{M}arkovianity in quantum and classical
  systems.
\newblock {\em New Journal of Physics}, 13(9):093004, 2011.

\bibitem{guarnieri2014quantum}
Giacomo Guarnieri, Andrea Smirne, and Bassano Vacchini.
\newblock Quantum regression theorem and non-{M}arkovianity of quantum
  dynamics.
\newblock {\em Physical Review A}, 90(2):022110, 2014.

\bibitem{VA17}
In\'es de~Vega and Daniel Alonso.
\newblock Dynamics of non-{M}arkovian open quantum systems.
\newblock {\em Rev. Mod. Phys.}, 89:015001, Jan 2017.

\bibitem{breuer2016colloquium}
Heinz-Peter Breuer, Elsi-Mari Laine, Jyrki Piilo, and Bassano Vacchini.
\newblock Colloquium: Non-{M}arkovian dynamics in open quantum systems.
\newblock {\em Rev. Mod. Phys}, 88(2):021002, 2016.

\bibitem{li2018concepts}
Li~Li, Michael~J.W. Hall, and Howard~M. Wiseman.
\newblock Concepts of quantum non-{M}arkovianity: A hierarchy.
\newblock {\em Physics Reports}, 759:1 -- 51, 2018.

\bibitem{RHP10}
{\'A}ngel Rivas, Susana~F Huelga, and Martin~B Plenio.
\newblock Entanglement and non-{M}arkovianity of quantum evolutions.
\newblock {\em Phys. Rev. Lett}, 105(5):050403, 2010.

\bibitem{RHP14}
Angel Rivas, Susana~F Huelga, and Martin~B Plenio.
\newblock Quantum non-{M}arkovianity: characterization, quantification and
  detection.
\newblock {\em Rep. Prog. Phys}, 77(9):094001, 2014.

\bibitem{breuer2009measure}
Heinz-Peter Breuer, Elsi-Mari Laine, and Jyrki Piilo.
\newblock Measure for the degree of non-{M}arkovian behavior of quantum
  processes in open systems.
\newblock {\em Phys. Rev. Lett}, 103(21):210401, 2009.

\bibitem{chruscinski2014degree}
Dariusz Chru{\'s}ci{\'n}ski and Sabrina Maniscalco.
\newblock Degree of non-{M}arkovianity of quantum evolution.
\newblock {\em Physical review letters}, 112(12):120404, 2014.

\bibitem{chruscinski2017detecting}
Dariusz Chru{\'s}ci{\'n}ski, Chiara Macchiavello, and Sabrina Maniscalco.
\newblock Detecting non-{M}arkovianity of quantum evolution via spectra of
  dynamical maps.
\newblock {\em Physical review letters}, 118(8):080404, 2017.

\bibitem{strasberg2018response}
Philipp Strasberg and Massimiliano Esposito.
\newblock Response functions as quantifiers of non-{M}arkovianity.
\newblock {\em Physical review letters}, 121(4):040601, 2018.

\bibitem{pollock2018non}
Felix~A. Pollock, C\'esar Rodr\'{\i}guez-Rosario, Thomas Frauenheim, Mauro
  Paternostro, and Kavan Modi.
\newblock Non-{M}arkovian quantum processes: Complete framework and efficient
  characterization.
\newblock {\em Phys. Rev. A}, 97:012127, Jan 2018.

\bibitem{mazzola2010phenomenological}
Laura Mazzola, E-M Laine, H-P Breuer, Sabrina Maniscalco, and Jyrki Piilo.
\newblock Phenomenological memory-kernel master equations and time-dependent
  {M}arkovian processes.
\newblock {\em Physical Review A}, 81(6):062120, 2010.

\bibitem{li2019non}
C-F Li, G-C Guo, and J~Piilo.
\newblock Non-{M}arkovian quantum dynamics: What does it mean?
\newblock {\em EPL (Europhysics Letters)}, 127(5):50001, 2019.

\bibitem{li2020non}
C-F Li, G-C Guo, and J~Piilo.
\newblock Non-{M}arkovian quantum dynamics: What is it good for?
\newblock {\em EPL (Europhysics Letters)}, 128(3):30001, 2020.

\bibitem{lindblad1976}
G.~Lindblad.
\newblock {\em Commun. Math. Phys.}, 48:119, 1976.

\bibitem{breuer2002theory}
Heinz-Peter Breuer and Francesco Petruccione.
\newblock {\em The theory of open quantum systems}.
\newblock Oxford University Press, 2002.

\bibitem{banerjee2018open}
Subhashish Banerjee.
\newblock {\em Open quantum systems: Dynamics of Nonclassical Evolution}.
\newblock Springer Nature Singapore, 2018.

\bibitem{chruscinski2010non}
Dariusz Chru{\'s}ci{\'n}ski and Andrzej Kossakowski.
\newblock Non-{M}arkovian quantum dynamics: local versus nonlocal.
\newblock {\em Physical review letters}, 104(7):070406, 2010.

\bibitem{ollivier2001discord}
Harold Ollivier and Wojciech~H Zurek.
\newblock Quantum discord: a measure of the quantumness of correlations.
\newblock {\em Phys. Rev. Lett}, 88(1):017901, 2001.

\bibitem{rao2011quantumness}
Balaji~R Rao, R~Srikanth, CM~Chandrashekar, and Subhashish Banerjee.
\newblock Quantumness of noisy quantum walks: a comparison between
  measurement-induced disturbance and quantum discord.
\newblock {\em Phys. Rev. A}, 83(6):064302, 2011.

\bibitem{shrikant2018non-Markovian}
U.~Shrikant, R.~Srikanth, and Subhashish Banerjee.
\newblock Non-{M}arkovian dephasing and depolarizing channels.
\newblock {\em Phys. Rev. A}, 98:032328, Sep 2018.

\bibitem{chandrashekar2007symmetries}
CM~Chandrashekar, R~Srikanth, and Subhashish Banerjee.
\newblock Symmetries and noise in quantum walk.
\newblock {\em Phys. Rev. A}, 76(2):022316, 2007.

\bibitem{srikanth2008squeezed}
R~Srikanth and Subhashish Banerjee.
\newblock Squeezed generalized amplitude damping channel.
\newblock {\em Phys. Rev. A}, 77(1):012318, 2008.

\bibitem{omkar2013dissipative}
S~Omkar, R~Srikanth, and Subhashish Banerjee.
\newblock Dissipative and non-dissipative single-qubit channels: dynamics and
  geometry.
\newblock {\em Qu. Inf. Proc}, 12(12):3725--3744, 2013.

\bibitem{choi1975completely}
Man-Duen Choi.
\newblock Completely positive linear maps on complex matrices.
\newblock {\em Linear algebra and its applications}, 10(3):285--290, 1975.

\bibitem{kumar2017nonmarkovian}
N.~Pradeep Kumar, Subhashish Banerjee, R.~Srikanth, Vinayak Jagadish, and
  Francesco Petruccione.
\newblock Non-{M}arkovian evolution: a quantum walk perspective.
\newblock {\em Open systems \& Information Dynam}, 25(03):1850014, 2018.

\bibitem{sudarshan1961stochastic}
E.~C.~G. Sudarshan, P.~M. Mathews, and Jayaseetha Rau.
\newblock Stochastic dynamics of quantum-mechanical systems.
\newblock {\em Phys. Rev.}, 121:920--924, Feb 1961.

\bibitem{horodecki2009quantum}
Ryszard Horodecki, Pawe{\l} Horodecki, Micha{\l} Horodecki, and Karol
  Horodecki.
\newblock Quantum entanglement.
\newblock {\em Reviews of modern physics}, 81(2):865, 2009.

\bibitem{yu2010entanglement}
Ting Yu and J.H. Eberly.
\newblock Entanglement evolution in a non-{M}arkovian environment.
\newblock {\em Optics Communications}, 283(5):676 -- 680, 2010.

\bibitem{uhlenbeck1930theory}
George~E Uhlenbeck and Leonard~S Ornstein.
\newblock On the theory of the brownian motion.
\newblock {\em Physical review}, 36(5):823, 1930.

\bibitem{paladino20141}
E~Paladino, YM~Galperin, G~Falci, and BL~Altshuler.
\newblock 1/f noise: Implications for solid-state quantum information.
\newblock {\em Reviews of Modern Physics}, 86(2):361, 2014.

\bibitem{andersson2007finding}
Erika Andersson, Jim~D. Cresser, and Michael J.~W. Hall.
\newblock Finding the {K}raus decomposition from a master equation and vice
  versa.
\newblock {\em J. Mod. Opt}, 54:1695, 2007.

\bibitem{hall2014canonical}
Michael J.~W. Hall, James~D. Cresser, Li~Li, and Erika Andersson.
\newblock Canonical form of master equations and characterization of
  non-{M}arkovianity.
\newblock {\em Phys. Rev. A}, 89:042120, Apr 2014.

\bibitem{gorini1976}
V.~Gorini, A.~Kossakowski, and E.~C.~G. Sudarshan.
\newblock {\em J. Math. Phys.}, 17:821, 1976.

\bibitem{wolf2008assessing}
Michael~Marc Wolf, J~Eisert, TS~Cubitt, and J~Ignacio Cirac.
\newblock Assessing non-{M}arkovian quantum dynamics.
\newblock {\em Physical review letters}, 101(15):150402, 2008.

\bibitem{daffer2004depolarizing}
Sonja Daffer, Krzysztof W{\'o}dkiewicz, James~D Cresser, and John~K McIver.
\newblock Depolarizing channel as a completely positive map with memory.
\newblock {\em Phys. Rev. A}, 70(1):010304, 2004.

\bibitem{shrikant2020quasi}
U~Shrikant, Vinod~N Rao, R~Srikanth, and Subhashish Banerjee.
\newblock Quasi-eternally non-{M}arkovian channels with a multiply singular
  generator.
\newblock {\em arXiv preprint arXiv:2002.11452}, 2020.

\bibitem{budini2018maximally}
Adri{\'a}n~A Budini.
\newblock Maximally non-{M}arkovian quantum dynamics without
  environment-to-system backflow of information.
\newblock {\em Physical Review A}, 97(5):052133, 2018.

\bibitem{swain1981master}
S~Swain.
\newblock Master equation derivation of quantum regression theorem.
\newblock {\em Journal of Physics A: Mathematical and General}, 14(10):2577,
  1981.

\bibitem{bhattacharya2018resource}
Samyadeb Bhattacharya, Bihalan Bhattacharya, and A.~S Majumdar.
\newblock Resource theory of non-{M}arkovianity: A thermodynamic perspective.
\newblock {\em arXiv preprint arXiv:1803.06881}, 2018.

\end{thebibliography}

\end{document}